\documentstyle[12pt]{article}

\title{
ONE-DIMENSIONAL FERMI  LIQUID and SYMMETRY BREAKING in the VORTEX CORE }
\author{Yu.G. Makhlin and  G.E. Volovik\\
Low Temperature Laboratory\\
Helsinki University of Technology\\
Otakaari 3A, 02150 Espoo, Finland\\
and\\
L.D. Landau Institute for Theoretical Physics, \\
Kosygin Str. 2, 117940 Moscow, Russia\\
}
\begin{document}
\maketitle
\begin{abstract}
{
Fermions localized within vortex cores  can form one-dimensional Fermi
liquids. The nonzero density of states in these Fermi-liquids can
lead to instability of the symmetric structure of the vortex core.  We
consider a symmetry breaking which is obtained due to  spontaneous
admixture of the spin-triplet $p$-wave component of the  order parameter in
conventional $s$-wave vortex in the presence of magnetic field. This occurs
at low enough temperature and leads to an asymmetric shape of the vortex core.
Similar phenomenon of the symmetry breaking induced by the core fermions occurs
in electroweak  $Z$-strings \cite{Naculich}. }
\end{abstract}

{\bf Introduction.}
Recently new techniques  have been developed, that in principle allow  to probe
the core structure of the individual Abrikosov vortex in superconductors. This
includes a scanning tunneling microscopy \cite{Hess}, electron holography,
Lorentz microscopy, etc. (see \cite{Tonomura} and Refs. there). That is why the
theoretical investigation of the symmetry of the core and of the energy
spectrum
of the fermions localized in the core is now up-to-date.

Three possible types of the energy spectrum $E(Q,k_z)$ of the fermions
localized
within the vortex core have been discussed in Ref.\cite{MisVol}. (i)  The
spectrum $E(Q,k_z)$, where $Q$ is the generalized angular momentum, has a gap
on
 order  $\Delta^2/E_F$ \cite{Caroli} (here $\Delta$ is the superconducting
gap and  $E_F$ is the Fermi energy in the normal Fermi liquid). (ii) One or
more
branches of the spectrum with particular $Q$'s cross the zero level as
functions
of the linear momentum $k_z$ at some points $k_{z}=k_{F,Q}$. In this case
fermions occupying negative energy levels form one-dimensional (1D)  Fermi
liquids with the Fermi points at $k_{F,Q}$ (see also an example of the
continuous vortex in $^3$He-A \cite{AphaseVortFermions}.  (iii) The flat band
where  all fermions have exactly zero energy in the finite region of the
momenta $k_z$, which is called the Fermi condensate \cite{Khodel}.

It was emphasized that  both the 1D Fermi liquid and  1D Fermi
condensate can be unstable towards a further breaking of the vortex symmetry
at
low temperatures, in particular the vortex can become nonaxisymmetric or the
vortex line can transform into a spiral \cite{MisVol} (see also
\cite{BphaseVortFermions}).

Recently such kind of instability of the vortex caused by the fermion zero
modes
in the core has been discussed for the $Z$-string solution in the
Weinberg-Salam model of electroweak interactions \cite{Naculich}. In this case
the situation corresponds to the case of 1D  Fermi liquid.  While in the
Ref.\cite{Naculich} it was stated that this can lead to the absolute
instability
of the $Z$-string, the situation is actually not dramatic for the existence of
the locally stable $Z$-string and is similar to that which has first
been discussed by Peierls \cite{Peierls}. In the Peierls model the 1D electron
liquid coupled to the lattice is unstable at low temperature towards the
dimerization of the lattice. In the $Z$-string the 1D massless relativistic
fermions coupled to the bosonic field of the order parameter (Higgs field) lead
below some critical temperature to the spontaneous symmetry breaking, which is
realized, e.g., in the development of the small amount of the upper component
of
the Higgs field or of another mode of instability. The transition temperature
$T_{c~{\rm core}}$ of such instability is usually exponentially small.

We discuss here the similar instability for the condensed matter vortices. We
have found that in most cases the transition temperature $T_{c~{\rm core}}$
is extremely small of the order $ \Delta \exp(-E_F/\Delta)$, but in some
cases  $T_{c~{\rm core}}$ can be reasonable, of the order $ \Delta^2/E_F$.

{\bf $s$-wave superfluids.}
Let us consider as an example the situation in superconductors with s-pairing.
We suppose that the particles are chargeless, and therefore we neglect all
magnetic effects. In the case the dependence of the order parameter on
coordinates around the axis of one-quantum vortex is given by $\Delta(r)\cdot
\exp(i\phi)$ in cylindrical coordinates. The magnitude of the gap $\Delta(r)$
equals zero at the axis $r=0$ and tends to the bulk value $\Delta$ for large
distances $r\gg\xi$, where $\xi=v_F/\Delta$ is the coherence length.

The system is invariant under translations along the axis $z$. If the
spin-orbit interaction is neglected there is also the $SU(2)_S$ group of spin
rotation, which is reduced to $SO(2)_S$ subgroup if the magnetic field is
applied. The "axial" symmetry of the order parameter in the vortex is
characterized by another $SO(2)_Q$ with a generator which is the combination of
the orbital momentum ${\bf L}_z$ and the particle number operator
${\bf I}$ \cite{VortexReview}:
\begin{equation}
{\bf Q}={\bf L}_z-\frac{m}{2}{\bf I}
\end{equation}
where $m=1$ is the winding number of the vortex. The fermion eigenstates are
characterized by the quantum numbers $k_z$, $Q$ and the direction of the spin
($\uparrow$ or $\downarrow$). The energy spectrum of low-lying excitations is
given by
\cite{Caroli}:
\begin{equation}
E(Q,k_z)=Q\cdot\omega_0(k_z)   \label{energy}
\end{equation}
with $\omega_0\sim\Delta^2/E_F$ for $k_z\ll
 k_F$ and $\omega_0\to\infty$
for $k_z\to k_F$ (this approximation is true for $\omega_0\ll\Delta$)
\cite{Caroli}. For $m=1$-quantum vortex $Q$ can assume
only half-integer values ($L_z$ is integer, and $I$ is $+1$ and $-1$ for
particles and   holes correspondingly).

Let us apply a magnetic field $H$ parallel to the axis of the vortex. Then
the energy spectrum is shifted down (up) for fermions with spin up (down):
\begin{equation}
E_{\uparrow,\downarrow}(Q,k_z)=Q\cdot\omega_0(k_z)\mp\mu H,
\end{equation}
and we get the picture (fig.\ref{ZeroModes}) where the branches
$E_{Q,\uparrow}(k_z)$ and $E_{Q,\downarrow}(k_z)$ intersect for $\mu
H>Q\omega_0(0)/2$.

The fermions near the points $k_{F,Q}$ can be considered as a
one-dimensional Fermi-liquid, and $k_{F,Q}$ plays the part of the
Fermi-momentum. The nonzero density of states can lead to a new pairing of the
one-dimensional fermions, the character of which depends on the sign and the
magnitude of the interaction of 1D fermions in different channels. Let us show
that both symmetries $SO(2)_S$ and  $SO(2)_Q$ will be broken at low enough
temperatures, if the interaction in the spin-triplet $p$-wave channel is
attractive,and the core of the vortex can become anisotropic.

{\it $p$-wave component in the core}. In our model an extra interaction of the
fermions in 1D Fermi liquid results from the interaction of the initial bare
fermions in the $p$-channel and we are looking for the instability of the
vortex
core towards nucleation of the admixture of the spin triplet $p$-wave component
in the core of the type:
\begin{equation}
\delta\hat\Delta=g~\Delta
({\bf d}\cdot\hat{\vec\sigma})i\hat\sigma_2 ({\bf a}\cdot{\bf k}) ~ f(r)
\exp(iN\phi)~.
\label{dD}
\end{equation}
As in Ref.\cite{Naculich} we assume that $f(r)$ is concentrated in the core
since
it should influence only the core fermions. $f(r)$  is of the order of unity in
the core and decreases to zero outside the core of the vortex (i.e., at
$r\sim\xi$); in addition  $f(0)=0$, if $N\neq 0$. The parameter $\Delta$ is the
bulk value of the $s$-wave gap and  the dimensionless parameter $g$ is thus
the
magnitude of the admixture of the $p$-wave order parameter in the core region
as
compared to the $s$-wave component. For simplicity we can assume a trial
function of the form $f(r)= \exp(-r/\xi)$ for $N=0$. The spin vector ${\bf d}$
should be orthogonal to the magnetic field to break the spin rotation symmetry.

The suggested change of the order parameter breaks ${\bf Q}$-symmetry as well.
To show this note that $\delta\hat\Delta$ is the sum of three orbital
harmonics: ${\bf a}\cdot{\bf k}=a_+k_- + a_-k_+ + a_0k_z$ where $k_\pm=k_x\pm
ik_y$. We suppose that only one $a$ is nonzero, namely $a_{l}$ with some
value $l=0,\pm 1$ of the $z$-projection of the Cooper pair orbital momentum.
It is easy to check that new order parameter is not ${\bf Q}$-invariant for
$N+l\ne 1$ and this means the violation of the axial symmetry of the whole
core structure. $g$ is the dimensionless parameter which shows the strength of
the deformation. This symmetry breaking leads to mixing of $Q,\uparrow$ and
$-Q,\downarrow$ fermions states in the same way as $u$ and $d$ quarks are mixed
in the $Z$-string in the Naculich scenario \cite{Naculich} of the instability
of
$Z$-string. Due to the noncrossing theorem for the levels of the same symmetry,
the two branches which crossed each other at $k_{F,Q}$ repel each other and the
gap between the states appears which is proportional to $g$.

{\it Deformation of the fermionic spectrum}. Without the admixture the
Bogolyubov-Nambu hamiltonian is given by $4\times 4$ matrix
\begin{equation}
{\cal H}_0=\left(
\begin{array}{cccc}
\xi&0&0&\Delta\\
0&\xi&-\Delta&0\\
0&-\Delta^*&-\xi& 0\\
\Delta^*&0&0&-\xi
\end{array}
\right)
{}.
\end{equation}
Here $\xi=(1/2m)(-d^2/d{\bf r}^2-k_F^2)$. The eigenstates of ${\cal H}_0$
with energies (\ref{energy})
for spin up and distances $r\gg k_F^{-1}$ are given by \cite{Caroli}
\begin{equation}
\chi_{k_z,Q,\uparrow}=
\left(
\begin{array}{c}
\chi_1\\ 0\\ 0\\ \chi_2
\end{array}
\right)=
const\cdot e^{ik_zz}e^{iQ\phi} H_\nu (qr) e^{-K(r)}
\cdot\left(
\begin{array}{c}
\exp\left(\frac{i}{2}(\phi+\psi(r))\right)\\ 0\\ 0\\
-i\exp\left(-\frac{i}{2}(\phi+\psi(r))\right)
\end{array}
\right)
\end{equation}
where $H_\nu$ is the Hankel function with index $\nu=\sqrt{Q^2+1/4}$;
$q^2=k_F^2-k_z^2$,
$$
K(r)=\frac{m}{q}\int\limits_0^r\Delta(r')dr',
$$
$$
\psi(r)=-\int\limits_r^\infty\exp(2(K(r)-K(r')))
\left(\frac{2Em}{q}+\frac{Q}{qr'^2}\right)dr'.
$$
For spin down the eigenstates are
\begin{equation}
\chi_{k_z,Q,\downarrow}=
\left(
\begin{array}{c}
0\\ \chi_1\\ -\chi_2\\ 0
\end{array}
\right).
\end{equation}

The change $\delta\hat\Delta$ leads to the perturbation
of the mean field Bogolyubov-Nambu hamiltonian.
The angular integral in the matrix element of
this perturbation between the states $\chi_{k_z,Q,\uparrow}$ and $\chi_{k_z,-Q,
\downarrow}$ does not vanish only for
\begin{equation}
N+l+1 \pm 2Q=0 \label{NofQ}~,
\end{equation}
(i.e. $N+l$ should be an even integer). In this case the matrix element
between the
spin-up and spin-down states is of the order of $g\Delta$. For other $Q$ the
integral over $d\phi$ in the matrix element vanishes. The $2\times 2$
hamiltonian for two mixing states
$(k_z,Q,\uparrow)$ and
$(k_z,-Q,\downarrow)$ for $k_z$ near $k_{F,Q}$ is then given by:
\begin{equation}
\left(
\begin{array}{cc}
E_{old} & g\Delta\\
g\Delta & -E_{old}
\end{array}
\right) .
\end{equation}
Here $E_{old}= v_{F,Q}(k_z-k_{F,Q})$, the derivative
$v_{F,Q}=Q\omega_0^\prime$ of the energy over momentum $k_z$ represents the
"Fermi" velocity of the 1D Fermi-liquid. The modified energy spectrum takes the
form
\begin{equation}
E_{new}=\pm\sqrt{v_{F,Q}^2(k_z-k_{F,Q})^2+
g^2\Delta^2}.
\end{equation}
At zero temperature fermions occupying the states with negative energy near
$k_{F,Q}$ gain in energy after the change in the order parameter. We are
going now to compare energy gain and energy losses due to the perturbation $g$.

{\it Energy balance and zero-temperature deformation of the vortex core.}
The main contribution to the difference  between the energy  of the fermionic
vacuum in symmetric and nonsymmetric states comes from the logarithmic term
which arises due to  appearance of the energy gap in 1D Fermi-liquid:
\begin{eqnarray}
E_{gain}&=&\int\frac{dk_z}{2\pi} (E_{new}-E_{old}) \nonumber\\
&\cong& -\frac{g^2\Delta^2}{v_{F,Q}}
\int\limits_{k_1}^{k_2}
\frac{d(k-k_{F,Q})}{|k-k_{F,Q}|}~. \label{Egain}
\end{eqnarray}
The lower cut-off is determined by the gap itself
\begin{equation}
k_1\cong \frac{g\Delta} {v_{F,Q}}~.
\end{equation}
The most interesting situation takes place for $k_{F,Q}$   not too close to
$k_F$, in this case the Fermi-velocity $v_{F,Q}$ is small and the energy
gain increases. This occurs in  field  $H$ near its threshold values $H_c(Q)$
at
which the branches $Q,\uparrow$ and $-Q,\downarrow$ start to cross each other
and the ``Fermi-momentum'' $k_{F,Q}$ appears for the first time. The
``Fermi-momentum'' $k_{F,Q}$ is given by the condition $Q\omega_0(k_{F,Q})=\mu
H$, that is we are interested in the fields $\mu H\sim Q\Delta^2/E_F$. In this
region the upper cut-off in the logarithmically divergent integral in
(\ref{Egain}) is $k_2\sim k_{F,Q}\sim k_F\sqrt{ {\mu H\over Q\omega_0(0)}-1}$.
This leads to the result
\begin{equation}
E_{gain}\cong E_Fk_F g^2 \frac{k_F}{|Q|k_{F,Q}}
\ln\left(\frac{|Q|\Delta k_{F,Q}^2}{gE_Fk_F^2}\right).
\end{equation}

On the other hand the energy loss due to the perturbation of the
superconducting
order parameter in the core of the vortex is of the order of $E_{loss}\cong
E_Fk_Fg^2/\gamma_1$. Here $\gamma_1$ is the relative magnitude of the
attractive interaction of bare fermions in the $p$-channel compared to that in
$s$-channel. For
$N\ne 0$ the contribution to the gradient energy of space inhomogeneity of
$\delta\Delta$ increases by the amount
$\sim E_Fk_FN^2g^2$. This consideration is valid only if the latter  term is
smaller than the former one.

Comparison of the energies shows the instability,
the relative magnitude $g$ of the new $p$-wave order parameter at zero
temperature  being of the order of
\begin{equation}
g\cong \frac{\Delta}{E_F} |Q| \left(\frac{k_{F,Q}}{k_F}\right)^2
\exp\left(-|Q| \frac{k_{F,Q}}{k_F} ({1\over \gamma_1}+N^2) \right)~.
\label{g}
\end{equation}
Here we omitted the coefficients of the order of unity.
$N(Q)$ is given by (\ref{NofQ}). At small $Q$ the exponent
$\exp(-O(1))$ is not too small for $k_{F,Q}\sim k_F$, and the transition
temperature is reasonable. On the other hand the numerical factor in the
exponent $O(1)$ can be large. In this case (even for large Q) we can take
$k_{F,Q}/k_F$ to be small which leads to the exponential increase in the
magnitude of $g$ followed by just power-law decrease in the preexponential
factor.

{\it Symmetry breaking scheme.}
The above result means that there should be a
second-order phase transition at $T_{c ~{\rm core}}$ into the broken symmetry
state. This is the state with the additional superfluid order parameter
(\ref{dD}) with $p$-pairing. The $\bf k$-dependence of this order parameter is
described by the quantum number  $l=0,\pm 1$, and the spatial
$\phi$-dependence by another integer $N=\pm 2Q-l-1$. So, there are six
competing structures (for three values of $l$) with comparable transition
temperatures. They correspond to different symmetry groups of the new
vortex core structure below $T_{c~{\rm core}}$.

The spin structure of the new order parameter is described by the vector
$\bf d$ in the $xy$-plane. For ${\bf d}=\hat x\pm i\hat y$ the symmetry
$SO(2)_Q\times SO(2)_S$ is reduced to the combined symmetry group
$SO(2)_{{\bf Q}\pm 2|Q\pm 1|\bf S}$. For other $\bf d$ in the plane the
symmetry is reduced to the discrete group $Z_{4|Q\pm 1|}$ the elements of which
are $\bf Q$-rotations accompanied by the possible inversion of $\bf d$.
In the former case, say, density is still axially symmetric, and in the latter
case it is not.

{\it Transition temperatures}.
The  temperature $T_{c ~{\rm core}}$ at which the transition to the asymmetric
core state occurs is
\begin{equation}
T_{c ~{\rm core}}\sim g\Delta~.
\label{Tc}
\end{equation}
(see the phase diagram, fig.\ref{TransTemp}). The maximal transition
temperature in the $Q$-th region and the range of the fields where the bare
structure is unstable (the height and the width of the region in $H-T$-plane)
decrease with $Q$:
\begin{equation}
T_{c,max}(Q)\sim \Delta\mu H(Q)\sim
\frac{\Delta^2}{E_F}\frac{1}{Q^5}.
\end{equation}
The minimal field is of order $(\Delta^2/E_F)/\mu$. This field can be less than
the upper critical field $H_{c2}$ if the effective fermion mass $m^\star$ is
larger than the bare electron mass $m$.

{\bf Vortex in $^3$He-B.}
In the spin-triplet superfluid $^3$He-B  the
situation is almost equivalent to that in s-superfluids. In  the most symmetric
B-phase vortex the branches of the energy spectrum are described by quantum
numbers $k_z$, $Q$ and helicity $\lambda=\pm 1$ \cite{MisVol}. The branches
$E_{Q=0,\lambda=1}(k_z)$ and $E_{Q=0,\lambda=-1}(k_z)$ intersect at the
the point $k_z=0$, $E=0$ even in zero external magnetic field.  The branches
$E_{Q,\lambda=1}$ and $E_{-Q,\lambda=-1}$ begin to intersect at $k_z\sim k_F$
in the fields $\mu H_c(Q)\sim Q\omega_0\sim Q\Delta^2/E_F$ (fig.\ref{Bphase}).

In the case there is no conservation of longitudinal spin component $S_z$,
and the perturbing order parameter may have vector ${\bf d}$ of spin
anisotropy parallel to $z$-axis. The deformation of the vortex can
occur within the triplet pairing state; the singlet pairing
(e.g., s-pairing) is accepted as well.

Transition temperatures, critical fields and the whole phase diagram are
more or less the same as for s-superfluids. For the two branches mentioned
above which intersect in zero field the transition temperature is of the order
of $T_c\sim \Delta^2/E_F$. It was found
experimentally and theoretically that the most symmetric
vortex in $^3$He-B is unstable towards the symmetry breaking at least at high
temperature of order
$T_c$ \cite{VortexReview,Fog,BPhase vortex}. Thus if the symmetry is restored
at
lower temperature it is broken again at very low $T$.

The states with definite $\lambda$ are the mixtures of states with different
particle number and different $S_z=\pm 1/2$. The analysis shows that in the
$Q$-th region the angle dependence of the perturbation of the order parameter
can be given by an integer
$$
N=\pm 2Q-l-m+\Sigma
$$
where $\Sigma=0,\pm 2$. Therefore for each $l$ we get three different
values of $N$ and thus several competing types of the symmetry breaking.

{\bf Conclusion.}
The vortex structure, which contains the one-dimensional Fermi-liquid
of the core fermions, is unstable towards the symmetry breaking. In the case
of Abrikosov  vortex in the $s$-wave superconductors this leads to the breaking
of the axial symmetry of the vortex core in some regions in $H-T$-plane.

Similar symmetry breaking occurs in the electroweak $Z$-string, where the
mixing of the $d$ and $u$ quarks opens the gap in the fermionic spectrum
\cite{Naculich}. In the Abrikosov and the $^3$He-B vortices this corresponds to
the hybridization of $Q,\uparrow$ and $-Q,\downarrow$ states of fermions. For
the Abrikosov vortex  this leads to appearance of the $p$-wave pairing
component
in the core.

We thank T. Vachaspati for valuable discussion.  This work was
supported through the ROTA co-operation plan of the Finnish Academy and the
Russian Academy of Sciences. G.E.V. was also supported  by the Russian
Foundation for Fundamental Sciences, Grant Nos. 93-02-02687 and 94-02-03121.
Yu.G.M. was also supported by the International Science Foundation
and the Russian Government, Grant No.MGI300, and by the ``Soros
Post-Graduate Student'' program of the Open Society Institute.

\begin{figure}[h]\caption[ZeroModes]{Spectrum of fermions in the  vortex
in $s$-wave superconductor. The levels are shifted by magnetic field. Some
branches cross the zero energy level forming 1D Fermi liquid. The fermions
with opposite spins on the Fermi level are paired at low $T$ which leads to
broken symmetry of the core.}
\label{ZeroModes}
\end{figure}

\begin{figure}[h]\caption[TransTemp]{Schematic dependence of the temperature of
the core transtion on the  magnetic field. Within the shaded area the
symmetry of the vortex core is broken.}
\label{TransTemp}
\end{figure}

\begin{figure}[h]\caption[Bphase]{Spectrum of fermions in the most symmetric
vortex in $^3$He-B \protect\cite{MisVol}. $\lambda=\pm 1$ is helicity.}
\label{Bphase}
\end{figure}

\end{document}